\documentclass[aps,twocolumn]{revtex4}
\usepackage{graphicx}
\begin{document}

\title{Competing interactions in two dimensional Coulomb systems:
Surface charge heterogeneities in co-assembled cationic-anionic
incompatible mixtures}
\author{Sharon M. Loverde}
\author{Yury S. Velichko} \author{Monica Olvera de la
Cruz}\affiliation{Department of Materials Science and
Engineering\\Northwestern University, Evanston, Illinois
60208-3108}
\date{\today}

\begin{abstract}
A binary mixture of oppositely charged components confined to a
plane such as cationic and anionic lipid bilayers may exhibit
local segregation. The relative strength of the net short range
interactions, which favors macroscopic segregation, and the long
range electrostatic interactions, which favors mixing, determines
the length scale of the finite size or microphase segregation. The
free energy of the system can be examined analytically in two
separate regimes, when considering small density fluctuations at
high temperatures, and when considering the periodic ordering of
the system at low temperatures (F. J. Solis and M. Olvera de la
Cruz, \textit{J. Chem. Phys.} \textbf{122}, 054905 (2000)). A
simple Molecular Dynamics simulation of oppositely charged
monomers, interacting with a short range Lennard Jones potential
and confined to a two dimensional plane, is examined at different
strengths of short and long range interactions.  The system
exhibits well-defined domains that can be characterized by their
periodic length-scale as well as the orientational ordering of
their interfaces. By adding salt, the ordering of the domains
disappears and the mixture macroscopically phase segregates in
agreement with analytical predictions.
\end{abstract}

\maketitle

\section{Introduction}
Biological and synthetic heterogeneous charged molecules are
expected to self-organize in aqueous solutions into complex ionic
structures. Co-assemblies of oppositely charged molecules are
ubiquitous given that nucleic acids and most proteins are charged.
The structure of oppositely charged biomolecular co-assemblies
such as DNA-proteins in nucleosomes \cite{Livolant} and
actin-protein complexes in the cytoskeleton \cite{SandersPRL2005},
are the result of the competition of short range interactions,
including excluded volume, and electrostatics. Cationic and
anionic mixtures of lipids or peptide amphiphiles co-assembled
into vesicles \cite{Zemb,tirrell} or cylindrical micelles
\cite{Kaler,walker,Stupp} are examples of co-assemblies stabilized
by hydrophobic interactions and electrostatics. The surfaces of
such complexes of oppositely charged molecules may not be
homogenous if the chemically co-assembled structures have net
repulsive short range interactions among them, or if the charges
exposed to to surfaces have different degrees of compatibilities
with water.  Understanding the surface assembly of a complex group
of charged components may lead to a greater deal of understanding
concerning the stability of self-assembled aggregates, or
moreover, give insight into the complex behavior of lipid rafts
and their contribution towards protein sorting and cell signaling
\cite{liu2}.

Bulk solution properties of electrostatic driven co-assemblies of
cationic and anionic macroions have been extensively studied, such
as DNA in cationic molecules of valence $3+$ and higher
\cite{widom,raspaud}, as well as other synthetic strongly charged
polyelectrolytes in metallic multivalent salts \cite{cruz}. These
hydrated multivalent ions are known to induce the precipitation of
strongly charged chains of opposite charge into dense ionic
structures \cite{Levin,SolisOLD,Muthu}. Co-assemblies of
hydrophobic molecules of opposite charge, however, are less
understood. Surface heterogeneities in co-assembled chemically
incompatible oppositely charge molecules have been recently
predicted analytically \cite{Muthu}. The surface charge
heterogeneities are due to the competition between the long range
electrostatic interactions (which decay as $1/r$ because the
surface is embedded in a tree dimensional medium) and the short
range interactions. The net incompatibility among the chemically
different components of opposite charge promotes macroscopic
segregation. Electrostatic interactions, on the other hand,
promotes mixing into an ionic crystal structure. Consider cationic
molecules with strong attractions among themselves and restricted
to surfaces such as cationic lipids adsorbed onto the surface of
mica, which is negatively charged
\cite{ducker,IsraelachviliPNAS2005}. The cationic molecules will
aggregate into positively charged domains due to the strong net
van der Waals attraction among them. The size of the domain,
however, cannot grow past a characteristic size due to the high
energetic penalty associated with the creation of a charged
domain.  This results in ordered finite size domains on the
surface at low temperatures \cite{solis}.  Finite size charge
heterogeneities have been observed experimentally on charged
surfaces in the presence of adsorbed self-aggregating molecules of
oppositely charge \cite{IsraelachviliPNAS2005,SandersPRL2005}.
Moreover, lattice Monte Carlo simulations of incompatible cationic
and anionic molecules restricted to the surface of cylinders
reveal many interesting finite temperature effects as well as
various stripes structures along the cylinder at lower
temperatures \cite{VelichkoPRE2005}.

The formation of charged domains on a flat square lattice due to
the competition between Coulomb interactions and net short range
repulsion amongst oppositely charged molecules has been explored
by simulation at zero temperature \cite{loew} and also by mean
field arguments at high temperatures \cite{loew,VelichkoPRE2005}.
These stoichiometric mixtures develop ordered striped domains
possessing a characteristic width that depends on the strength of
the competing Coulomb and short range interactions at low
temperatures. At high temperatures percolated structures develop
that resemble a spinodal decomposition pattern during phase
segregation of binary systems, but growth is restricted, as in
block copolymer systems with microphase segregation
\cite{leibler2, lefebvre}. Here, we analyze the formation of the
charged domains in two dimensions via molecular dynamics
simulations at different ratios of the short and long range
interactions. We analyze the symmetric case of equal head size of
stoichiometric mixtures of +1 and -1 charges with different
effective interactions among them. Finite temperature effects are
discussed. In Section II analytic arguments are given for the
scaling of the surface charge domain size in various regimes of
the degree of incompatibility. We justify the existence of charged
domains in surfaces, as compared to a bulk three dimensional
system. In section III we describe our simulations. In Section IV
we discuss the results and the last section we give the
conclusions.

\section{Theory}

The phase behavior of the ionic mixture can be examined in a
simplistic manner analytically in two separate regimes. At higher
temperatures, we consider small density fluctuations around the
mean.  At low temperatures, when the system exhibits strongly
segregated domains, we assume the system is periodic. At low
temperature values, or high values of the magnitude of short range
attraction, the system exhibits well-defined periodic lamellar
when the charge surface coverage of the positive and negative
molecules are equal and confined to a flat surface. The free
energy of the system is dominated by the electrostatic cohesive
energy in addition to the interfacial contribution to the free
energy, which is characterized by the line tension, $\gamma$ per
thermal energy $k_{B} T$. Within the strong segregation regime,
the entropic contribution to the free energy is negligible.

Following the example of the free energy for a incompressible two
dimensional system of a mixture of positive and negative
components \cite{solis}, we generalize the results for a course
grained free energy scaling analysis for a $d$ dimensional system
of $N_{A}$ positively and $N_{B}$ negatively charged components
interacting with a three dimensional Coulombic $1/r$ potential.
The free energy can be written as sum of the total electrostatic
interactions and the contribution from the line tension of each
periodic segregated domain. Each charged domain is approximated by
a electroneutral unit cell which has a characteristic lattice
length $L$, dimensions $L^{d}$, and an associated charge density
$\sigma$. The net free energy per total number of particles $N =
N_{A} + N_{B}$, in units of $k_{B}T$, can be written as
\begin{equation}
\frac{F_{NET}}{N}=\frac{F_{cell}}{N_{cell}}\approx\frac{a^{d}}{L^{d}}\left(\gamma
s_{1}L^{d-1}+\frac{l_{B}\sigma^{2}
s_{2}(L^{d})^{2}}{L}\right)=\left(\frac{F_{o}a^{d}}{L_{o}^{d}}\right)F.
\end{equation}
Here, $s_{1}$ and $s_{2}$ are geometrical parameters that depend
on the characteristic geometry of the underlying unit cell,
$a^{d}$ represents the size of the particle, and $N_{cell}$
represents the number of particles per unit cell. The Bjerrum
length $l_{B}$ is given by,
\begin{equation} l_{B} = \frac{e^2}{4\pi \epsilon \epsilon_{r}
k_{B}T}.\end{equation} $F_{o}$ and $L_{o}$ are system dependent
parameters, defined by the minimization of the free energy of the
system with respect to the characteristic size of the system, $L$,
\begin{equation} F_{o} =
\left(\frac{\gamma^{2d-1}}{(l_{B}\sigma^{2})^{d-1}}\right)^{1/d}\end{equation}
and
\begin{equation} L_{o}=\left(\frac{\gamma}{l_{B}\sigma^{2}}\right)^{1/d}.\end{equation}
The dimensionless free energy per unit area in terms
of $s_{1}$ and $s_{2}$ is then described by
\begin{equation}
F = \frac{s_{1}}{D}+ s_{2}D^{d-1}
\end{equation}
where $D=L/L_{o}$ is the ratio of the characteristic size of the
unit lattice to the length of the system.  Minimizing the
dimensionless free energy with respect to $D$ gives the free
energy of the favored periodic structure as
\begin{equation}F=2\left((d-1)s_{2}s_{1}^{d-1}\right)^{1/d}\end{equation}
where
\begin{equation}D=\left(\frac{s_{1}}{(d-1)s_{2}}\right)^{1/d}.
\end{equation}
Depending on the area fraction of charge coverage, $f$, and the
geometry of the unit lattice cell, the free energy can be
calculated for different sets of crystalline structures. For an
ideally symmetric system, consisting of equal components of
positively and negatively charged molecules with similar head
group sizes, $f$ is $1/2$. The minimum free energy in this case,
for a two dimensional system, is characterized by lamellar
structures.

We consider a line tension that is proportional to the
immiscibility of the component molecules, $\chi$ . The
Flory-Huggins paramater, $\chi$, is defined as the difference in
the magnitudes of the short range interactions between two
components as $\chi =
\left(\epsilon_{12}-\frac{1}{2}(\epsilon_{11}+\epsilon_{22})\right)/k_{B}T$,
where the $\epsilon_{ij}$ represents the pair interaction energy
between $i$ and $j$. For a lower, or two dimensional system,
$L_{o}$, would be comparably larger than for a three dimensional
system due not only to the $1/d$ power law dependence but also to
the decreased value of the Bjerrum length $l_B$ for a surface in
contact with water. For a surface in contact with an aqueous
solution the mean permittivity of the medium is much higher than
in a dense three dimensional system, which decreases the Bjerrum
length $l_B$, and thus the magnitude of $L_{o}$, significantly.
For these reasons, patterning on a surface due to the competition
of electrostatic interactions with short range interactions, is
considerably more feasible than the creation of charge domains in
a bulk three dimensional system.

Comparing length scales with experimental systems, consider a two
dimensional system of a single layer of positively and negatively
charged lipids at an interface between water and an alternate
medium. The average dielectric permittivity of at the interface
$\epsilon_{i} \sim 40$, in between that of the water
$\epsilon_{water} \sim 80$ and that of the dense medium
$\epsilon_{medium} \sim 1$. This would correspond to a Bjerrum
length $l_{B} \sim 2 nm$ in terms of the a classical electrostatic
interaction between charged head groups of the lipids exposed to
the aqueous interface. Considering a large magnitude of the net
interaction between tails of interacting lipids at the interface
($\chi \sim 15 $), depending on the length of the hydrophobic tail
of the molecules ($\sim 20$ carbons) and the charge density of the
head-group ($\sim .6/nm^{2}$), this could correspond to a fairly
large equilibrium domain size $L_{o}$ ($\sim 80 nm$). Domains of
this size or larger have been seen for experimental systems of
competing short range and long range electrostatic interactions,
although in comparing with these systems, a variety of kinetic and
specific interaction effects should also be considered
\cite{kaler2,potemkin}.

Next, consider the opposite, high temperature regime. Since the
system does not exhibit well-defined periodic structures, the
entropic contribution to the free energy cannot be ignored.  In
this case, linear response theory or the Random Phase
Approximation for a compressible binary systems \cite{gonzalez2}
is used to describe the behavior of the correlations as a function
of the relative strength of the short range attraction and the
electrostatic interactions. The Random Phase Approximation
involves an expansion of the free energy of the system in terms of
density fluctuations, neglecting all terms of larger than second
order. For a general system of N components, where $i$ and $j$
represent components of a different type, the partition function
can be written as \cite{borue,cruz}
\begin{equation}Z=\frac{1}{N_{A}!N_{B}!}\int exp\left(-\frac{H(r_{i}^{(1)}r_{j}^{(2)})}{k_{B}T}\right)\prod_{i}dr_{i}^{(1)}\prod_{j}dr_{j}^{(2)}\end{equation}
where the Hamiltonian of the system is represented by
\begin{equation}H(r_{i}^{(1)}r_{j}^{(2)})=\sum_{i}\sum_{j}\upsilon_{ij}(r_{i}^{(1)}-r_{j}^{(2)}).\end{equation}
It is assumed that the interparticle potential can be broken up
into a short range and long range electrostatic potential,
$\upsilon_{ij}=\upsilon_{ij}^{SR}+\upsilon_{ij}^{el}$. The short
range contribution is assumed to be of the form of the Fourier
transform of a Gaussian potential, which has been shown to
reasonably predict thermodynamic properties of binary systems
\cite{prestipino}, \begin{equation} \upsilon_{ij}^{SR}(r) =
\frac{\epsilon_{ij}}{\pi a^{2}} e^{-r^{2}/a^{2}}.\end{equation}
The long range potential is represented by the Debye H\"{u}ckel
potential,\begin{equation}\upsilon_{ij}^{el}(r) =
\frac{z_{i}z_{j}l_{B}e^{-\kappa r}}{r}\end{equation} where
$\kappa$, the inverse screening length, is defined by the
concentration of salt in the solution.  We assume that the density
is a smooth function and can be represented by the sum of its
Fourier components
\begin{equation}\rho^{i}(r)=\sum_{k}\rho_{k}^{i}e^{ikr}.\end{equation}
In this case, the partition function becomes
\begin{eqnarray}\lefteqn{ Z =
Z_{o}\frac{A^{N_{A}}A^{N_{B}}}{N_{A}!N_{B}!}\times} \nonumber \\ &
& \int e^{\left(-\frac{1}{2A}\sum_{k\neq
0}\sum_{ij}(\textbf{U}_{k}^{ij}+\rho_{i}^{-1}\delta_{ij})\rho_{k}^{i}\rho_{-k}^{j}
\right)}\times \nonumber \\ & &
\prod_{k>0}\prod_{i}\frac{d\rho_{k}^{i}}{\pi V \rho_{i}}.
\end{eqnarray} where $A$ represents the area of a two
dimensional plane in a three dimensional volume $V$. $Z_{o}$
includes the $k$ zero and the self energy terms.
$\textbf{U}_{k}^{ij}$ is the sum of the interaction energies of
the system, consisting of the short range interactions due to the
excluded volume and hydrophobic interactions,
$\upsilon_{ij}^{SR}(k)$, as well as the long range electrostatic
potential, $\upsilon_{ij}^{el}(k)$.

For an incompressible system of $i$ same-sized components,
\begin{equation} \sum_{i}\rho^{i}_{k} = 0.\end{equation} For the case of a
incompressible, neutral, symmetric system we also assume that
$\rho_{+}(k)=-\rho_{-}(k)$.  The electrostatic potential is the
two dimensional Fourier transform of the screened Coulomb
interaction between charge density fluctuations,
\begin{equation}U_{el}(k)= \int d^{2}r  e^{ik\cdot r} \frac{\sigma z_{T}^{2}l_{B}e^{-\kappa r}}{r} = \frac{1}{2} \sigma z_{T}^{2} \frac{2 \pi l_{B}}{\sqrt{\kappa^{2}+k^{2}}}\end{equation}
where $\sigma$ represents the charge density of the system and
$z_{T}$ represents the total positive and negative charge of the
components. In this case, the inverse structure factor has the
following contributions,
\begin{eqnarray}\frac{1}{S_0(k)}=U_{k}+\rho^{-1}= \frac{1}{\rho}+\frac{1}{1-\rho} -
2\chi+\chi k^{2}+U_{el}(k).\end{eqnarray}

The structure function has a peak at the most probable wave
lengths, $k^{*}$. For the case when there is no screening the
location of the peak, $k^{*}$, scales with the Bjerrum length,
$\l_{B}$ and magnitude of short range attraction, $\epsilon$, as
$k^{*} \sim (\epsilon/l_{B})^{\frac{1}{d+1}}$.  The scaling of the
periodic order of the system changes at the transition temperature
from $-(1+d)$ at higher temperatures considering small density
fluctuations to $-d$ at lower temperatures (Eq. 4), which is
predicted using the previously described theory of strong
segregation.  For a two dimensional system, which is the subject
of interest, the scaling is predicted to change from $-1/3$ to
$-1/2$ as the temperature decreases.

At high temperatures, in the nearly isotropic state, the total
free energy of the system per unit volume, in units of $k_BT$, can
be written as
\begin{eqnarray}
\frac{\Delta F\left(\phi\right)}{A
k_{B}T}=\phi\ln\phi+\left(1-\phi\right)\ln\left(1-\phi\right)-
\chi\phi^2  + F_{ele}/(k_{B}T), \label{free}
\end{eqnarray}
where $A$ represents the two dimensional area of a plane and
$F_{ele}/(k_{B}T)$ represents the one loop corrections obtained by
integrating the charge density fluctuations \cite{borue}.


\section{Model and Simulation Details}

The model system is composed of a mix of $N_{+}$ positively and
$N_{-}$ negatively charged monomer units in a simulation box of
size $L^{3}$.  The molecules are confined to a two-dimensional
plane perpendicular to the Z axis, with periodic boundary
conditions in the X and Y directions.  Each monomer unit
represents a charged biological or polymeric molecule, that
interacts attractively with a like monomer via hydrophobic forces.
In this paper, only symmetric mixtures are considered, where the
charge and radius of the positively and negatively charged monomer
units are equivalent.  The total system is electroneutral.  We are
interested in the case where the two-dimensional layer exhibits
well-defined periodic patterns along the surface of the plane.
Fluctuations perpendicular to the monomer plane are restricted.

Constant N,V,T Molecular Dynamics simulations were performed using
Espresso, simulation code developed by the MPIP-Mainz group of
Polymer Theory and Simulation (http://www.espresso.mpg.de).  A
stochastic or Langevin thermostat is used, to ensure a constant
temperature, along with a Verlet algorithm to calculate particle
velocities at each timestep.  The unit of energy is $\epsilon$, of
length $\sigma$, and of mass $m$. Temperature is then defined in
terms of $\epsilon/k_{B}T$ and time in units of
$\sqrt{\sigma^{2}m/\epsilon}$. A full Coulomb potential is used
for calculations of charge-charge interactions.  The ELC
(Electostatic Layer Correction) method developed by Arnold et al.
to sum the electrostatic energy contribution to the free energy
\cite{arnold, joannis}. This method is a correction to the P3M
Ewald summation technique \cite{hockney}, in which the Fourier
transform of the electrostatic contribution to the energy is
summed using a mesh formulation. In addition to full
electrostatics, a case of a screened Debye-H\"{u}ckel interaction
is considered to look at the subsequent melting of the periodic
structures when the potential is screened. Table 1 summarizes the
interaction potentials between the positive and negative component
monomers in the system.  The potential between charges is the full
Coulomb potential,
\begin{equation}
U_{C-ELC} = \frac{l_{B}Tq_{1}q_{2}}{r}
\end{equation}
where $l_{B}$ represents the Bjerrum length of the system.
\begin{table}[t]
\caption{\label{tab:table1}Interparticle Potentials }
\begin{ruledtabular}
\begin{tabular}{ccc}
Interactions & + & - \\
\hline
+ & $U_{C-ELC} + U_{LJ}$ & $U_{C-ELC} + U_{HC}$\\
- & $U_{C-ELC} + U_{HC}$ & $U_{C-ELC} + U_{LJ}$\\
\end{tabular}
\end{ruledtabular}
\end{table}
For the present simulation results, $l_{B}$'s of $0.1 \sigma, 0.2
\sigma$ and $0.5 \sigma$ are considered.  Considering an average
dielectric permittivity of the medium ($\epsilon_{r} \sim 80$)
this corresponds to a fairly large headgroup size ($\sim 20 \AA$).
The short range interaction between like monomers is the classic
Lennard Jones potential,
\begin{eqnarray}
U_{LJ} = 4\epsilon
\left(\left(\frac{\sigma}{r}\right)^{12}-\left(\frac{\sigma}{r}\right)^6\right)&&
r<r_{c}
\end{eqnarray}
where $\sigma$ is the monomer radius, and the potential is cut at
a radius $r_{c}$ of $2.5 \sigma$. An additional term is also added
to the potential to keep the derivative continuous at $r_{c}$.
$U_{HC}$ is the same as $U_{LJ}$, with a cutoff radius, $r_{c}$,
of $2^{1/6} \sigma$, including only the repulsive part of the
potential, which represents the excluded volume of the molecule.

Initially, a fairly dense surface density, $\rho$, of $0.6$ was
considered, to compare with phase behavior predicted by strong
segregation theory, while remaining sufficiently far from the two
dimensional hard disc crystallization regime of approximately
$\rho = 0.89$ determined by previous MC and MD simulations
\cite{jaster}. This also allows sufficient diffusion for the
system to equilibrate.
\begin{equation}\rho = \frac{(N_{+} + N_{-})\pi \sigma^{2} }{4 L^{2}} \end{equation}
Phase behavior in comparison with theory at lower surface
densities is slightly more complex and will be discussed in a
later paper.  The majority of simulation results are presented for
a system size of 1000 charged monomers, while finite size effects
are explored by increasing the system size by a factor of 2.
Approximately $10^{6}$ MD steps are used to equilibrate the
system; the equilibration time grows increasingly longer at higher
values of the Bjerrum length.

\section{Discussion of Two Dimensional Phase Behavior}

At lower values of $\epsilon$, domains of positive and negative
component monomers appear in the system.  As the magnitude of
$\epsilon$ increases, the domains begin to increase in size in an
isotropic manner, forming a percolated structure. As the value of
$\epsilon$ further increases, the domains begin elongate and then
to orient into well-defined lamellar, breaking the symmetry of the
system. Increasing even further, the lamellar begin to widen.
Average internal energy and heat capacity per particle are
calculated at several values of Bjerrum length ($l_{B} = 0.1
\sigma, 0.2 \sigma, 0.5 \sigma$). At higher values of $l_{B}$,
electrostatics plays a more important role in the equilibrium
configuration of the system. The electrostatic repulsion between
like charged monomers increases. In order to minimize this
contribution to the free energy, the stripes become thinner.
\begin{figure}[t]
\begin{center}
\includegraphics[width=8.5cm]{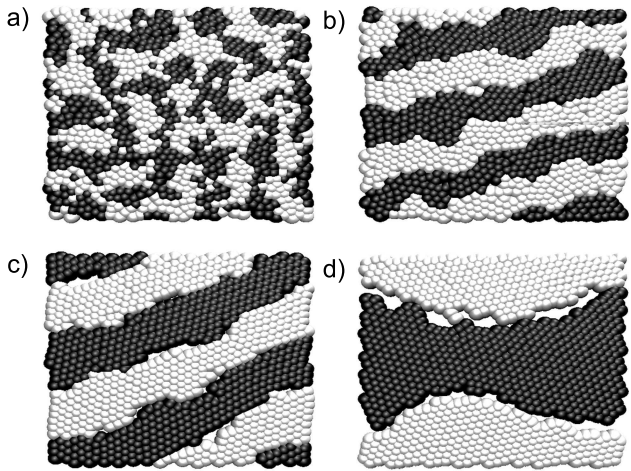}
\caption{\label{fig1} Snapshots of the system at $\epsilon = 1.0
(a), 2.5 (b), 4.0 (c) k_{B}T$ at a constant $l_{B}$ of $0.2
\sigma$. Introduction of $\kappa$ = 20$\sigma$ (d) induces
macroscopic phase segregation at an $\epsilon$ of $4.0 k_{B}T$,
$l_{B}$ of $0.2 \sigma$.}
\end{center}
\end{figure}
The average internal energy ($<E>/N$) and heat capacity per
particle ($C_{V}$) are calculated at two different values of the
Bjerrum length ($l_{B} = 0.1 \sigma, 0.2 \sigma, 0.5 \sigma$). The
average internal energy is less negative at the higher value of
$l_{B}$ due to the increased repulsion between like charged head
groups. At lower values of $l_{B}$ the heat capacity displays a
peak, which corresponds to the crossover from the percolated phase
to the lamellar phase. The magnitude of this peak increases and
shifts to the left as the value of the Bjerrum length is
decreased.
\begin{figure}[b]
\begin{center}
\includegraphics[width=8.5cm]{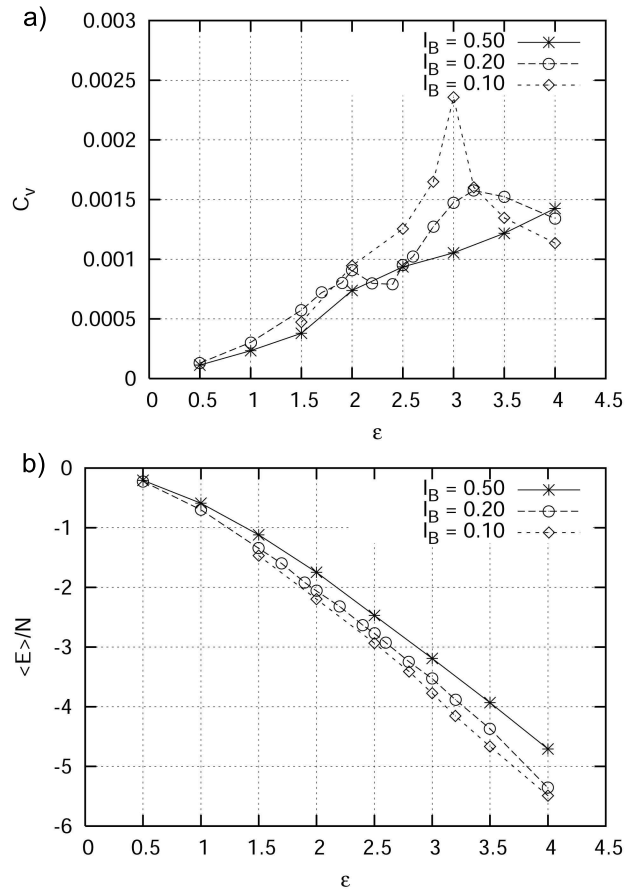}
\caption{\label{fig1} (a) Heat capacity per particle ($C_{V}$) and
(b) average internal energy ($<E>/N$) at several values of the
Bjerrum length ($l_{B} = 0.1 \sigma, 0.2 \sigma, 0.5 \sigma$). The
heat capacity displays a peak, which corresponds to a crossover
from percolated, random domains to a lamellar phase.  The
magnitude of the peak increases and shifts to the left as the
value of the $l_B$ is decreased.}
\end{center}
\end{figure}
Lamellar spacing is systematically characterized by the
calculation of the two dimensional structure factor, $S(\vec{k})$,
where $\vec{r}$ corresponds to a vector in the x,y plane.
\begin{eqnarray}
S(\vec{k}) = \int g(\vec{r}-\vec{r^{'}})e^{i\vec{k}\cdot
\vec{r}}e^{i\vec{k}\cdot \vec{r^{'}}}d^{2}\vec{r}
\end{eqnarray}
$S(\vec{k})$ displays a peak  at $k^{*}$ corresponding to the
inverse lamellar spacing in the direction perpendicular to the
lamellar. As a function of $\epsilon$, the peak location
corresponds to scaling predictions by strong segregation theory at
high values of $\epsilon$ ($k^{*}\sim \epsilon^{-1/2}$). At lower
values, the location is consistent with predictions by the Random
Phase Approximation ($k^{*}\sim \epsilon^{-1/3}$). The
orientational order of the domains can be characterized by the
interfacial orientational order parameter $g_{2}$
\cite{stoycheva},

\begin{eqnarray}
g_{2} =
\frac{1}{N}\sum^{N}_{i=1}\frac{1}{N_{i}}\sum^{N_{i}}_{j=1}e^{2i\theta_{ij}}
\end{eqnarray}
where $N_{i}$ is the number of neighbors of opposite type of
monomer at an interface and $\theta_{ij}$ is the angle between two
neighbors.  A neighbor is defined as two particles of different
type, within range of short range attraction ($r_{ij}<r_{c}$).  As
the magnitude of short range attraction increases, the calculated
order parameter increases, which corresponds to the ordering of
the domains by the development of orientational order at the
interface. The increase in order of the domains, indicated by an
increase in the order parameter $g_2$, proceeds the location of
the peak in the heat capacity.  Higher values of the Bjerrum
length $l_B$ correspond to a higher value of the order parameter
$g_2$ for stronger short range attraction.  As the electrostatic
contribution to the segregation increases, the characteristic
domain size decreases, but the orientational order of the domains
increases. At lower values of $l_{B}$, this initial increase is
followed by a levelling off, or slight decrease, that corresponds
to the formation of holes at the interface. The holes disrupt the
orientational order, or the hexagonal packing, of the monomers
within the segregated domains. This is equivalent to the inclusion
of a ternary component with a non-selective interaction between
positive and negative components.
\begin{figure}[t]
\begin{center}
\includegraphics[width=8.5cm]{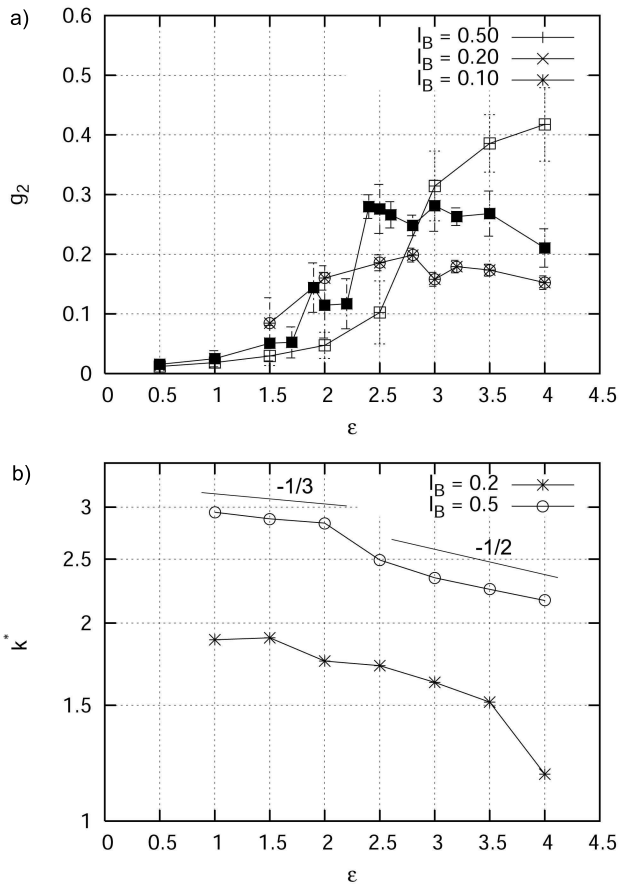}
\caption{\label{fig2} (a) The interfacial orientational order
parameter $g_{2}$ at several values of the Bjerrum length ($l_{B}
= 0.1 \sigma, 0.2 \sigma, 0.5 \sigma$) as a function of
$\epsilon$. As the orientational order of the domains increases,
$g_2$ increases from 0 to a finite value.(b) The location of the
peak $k^{*}$ in the structure factor $S(\vec{k})$ as a function of
$\epsilon.$  The scaling of $k^{*}$ with $\epsilon$ changes from
$-1/3$ to $-1/2$. }
\end{center}
\end{figure}
Initial examinations on the finite size effects of the system are
explored to determine the effect of the periodic boundary
conditions on the ordering of the more strongly segregated
lamellar.  Doubling the size of the system at larger values of the
short range attraction ($\epsilon = 4 k_{B}T$), quantitatively
affects the ordering of the lamellar by decreasing the alignment
of the domains along the boundaries of the system and increasing
the fluctuations along the interface.  This results in a
characteristic decrease in the order parameter, $g_{2}$ from
.43($\pm .02$) to .37($\pm .02$).  Further system sizes were not
examined due to sufficiently high surface density; the calculation
of the electrostatic energy is slow to converge at these density
ranges.

Introduction of electrostatic screening, or including the effects
of high salt on the local ordering of the system, is considered by
using a screened Debye H\"{u}ckel potential instead of the Coulomb
potential for electrostatic interactions,
\begin{equation}
U_{C-DH} = \frac{l_{B}Tq_{1}q_{2}e^{-\kappa r}}{r}
\end{equation}
where $\kappa$ represents the screening length due to the
surrounding three dimensional solution of ions (see Equation 10).
\begin{figure}[b]
\begin{center}
\includegraphics[width=8.5cm]{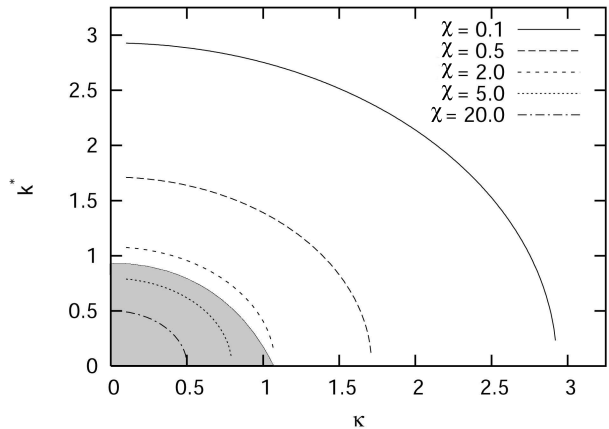}
\caption{\label{fig3} The location of the peak, $k^{*}$, in the
structure factor, $S(k)$, as a function of $\kappa$ as predicted
by linear response theory for several values of short range
attraction ($\chi = 0.1, 0.5, 2.0, 5.0, 20.0$) for an intermediate
strength of the electrostatics, $l_{B} = 0.2$.  The shaded area
indicates $S(k)$ diverges at a finite value of $k$.}
\end{center}
\end{figure}
At higher values of electrostatic screening ($\kappa = 5 \sigma,
10 \sigma, 15 \sigma$), examining the behavior of the system with
an intermediate value of short range attraction ($\epsilon =
2.0$), the system phase segregates into two macroscopic charged
domains of positive and negative ions. The peak in the structure
factor indicates that the segregation length-scale is nearly
constant as a function of the screening length.  This is in
agreement with analytic theory \cite{solis}. The location of the
peak shifts to lower values with an increase in the characteristic
size of the simulation box. To determine if these simulation
results are consistent with theoretical predictions, we examine
the behavior of the inverse structure factor (Eq. 16) in a regime
where the scaling of the peak in the structure factor from
simulation results is still consistent with linear response
theory.  We find that as $\kappa$, the magnitude of screening by
the ions of solution, increases, the value of $q^{*}$ goes
continuously to zero,
\begin{equation}q^{*} = \left( -\kappa^{2}+ (\frac{4 \pi
l_{B}}{\chi})^{2/3}\right)^{1/2}
\end{equation} before the structure factor diverges, at which
there is macroscopic phase segregation. At higher values of short
range attraction, the structure factor diverges when $q^{*}
> 0$.  This is in agreement with analytical predictions from the strong segregation regime, which predicts a discontinuous jump from finite sized periodic cells to an infinite cell at a value of $\kappa$ which is inversely proportional to periodic length-scale of the system \cite{solis}.

\begin{figure}[t]
\begin{center}
\includegraphics[width=8.5cm]{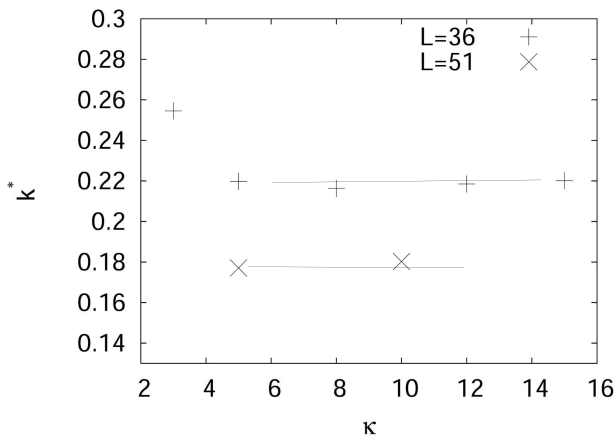}
\caption{\label{fig4} The location of the peak, $k^{*}$, in the
structure factor, $S(k)$, as a function of $\kappa$ from
simulation results spanning a range of the screening parameter
($\kappa = 3-15 \sigma$) at an intermediate value of short range
attraction ($\epsilon = 2.0$). Increasing the size of the system
from $L = 36$ to $L = 51$ decreases the average value of $k^{*}$.}
\end{center}
\end{figure}
\section{Conclusions}
Molecular Dynamics simulations of oppositely charged monomers,
interacting with a short range LJ potential and confined to a two
dimensional plane, are examined at different strengths of short
range attraction and long range electrostatics.  The system
exhibits well-defined domains; the width and ordering of the
domains are dependent on the depth of the LJ well, $\epsilon$, and
the strength of the Coulomb interactions, $l_{B}$.   The
length-scale of the ordering of the system can be quantitatively
characterized by the two dimensional Fourier transform of the
density, $S(\vec{k})$, where $\vec{k}$ is the inverse spacing of
the system. $S(\vec{k})$ has a peak $k^{*}$ which scales with the
line tension of the domains, $\gamma$. The underlying assumption
of strong segregation theory is that the microphase regions of
charge are well defined and periodic, with a line tension $\gamma$
that is proportional to $\chi$.  Since the magnitude of the short
range attraction $\epsilon$ is proportional the Flory-Huggins
interaction parameter $\chi$, $k^{*}$ should scale with $\epsilon$
in the regimes where strong segregation theory holds \cite{solis}.
It is shown that, at higher values of $\epsilon$, the scaling of
$k^{*}$ with $\epsilon$ is consistent with theory.  At lower
values of $\epsilon$, a different scaling is found, which is
consistent with that which is found using linear response theory.
Electrostatics represents a more important contribution to the
characterization of the interfacial line tension in this regime.

The degree of ordering can be examined by the calculation of the
interfacial orientational order parameter, $g_{2}$.  The
transition from a random, percolated domain structure to well
defined lamellar is a gradual transition, that is demonstrated by
the gradual increase of the parameter $g_{2}$ as a function of
$\epsilon$. This result is consistent with what one would exhibit
with a Kosterlitz and Thouless type transition \cite{jaster}, in
which the two dimensional ordering the system exhibits a
continuous phase transition, that can be defined by a similar
positional order parameter. Initial examinations of the finite
size effects of the system indicate that the degree of ordering is
slightly influenced by the periodicity of the simulation box,
however, further examinations were not made due to the
computational intensiveness of the electrostatic energy term.
Decreasing the strength of the electrostatics in the system, by
changing the charged interaction from a straight Coulomb potential
to a screened Debye H\"{u}ckel interaction, the ordering of the
system disappears and the mixture phase segregates, which is
consistent with analytical arguments.
\begin{acknowledgments}

This work was supported by the IGERT-NSF Fellowship awarded to S.
Loverde, and by NSF grant number DMR-0414446 and DMR-0503943.  S.
Loverde would like to acknowledge the group of Christian Holm, who
is now at Frankfurt Institute for Advanced Studies, Johann
Wolfgang Goethe-Universit\"{a}t, including Axel Arnold and
Bernward Mann for scientific and simulation discussions.

\end{acknowledgments}
\bibliographystyle{unsrt}
\bibliography{biblio}

\end{document}